\documentclass{aastex631}

\begin{document}

\title{Chemical Components in the Virgo Overdensity and Hercules-Aquila Cloud: hints of more than one merger event in GSE-like debris}
\shorttitle{Chemical Components in the VOD and HAC }
\shortauthors{Liu et al.}

\correspondingauthor{Cuihua Du}
\email{ducuihua@ucas.ac.cn}

\author{Haoyang Liu}
\affiliation{College of Astronomy and Space Sciences, University of Chinese Academy of Sciences, Beijing 100049, P.R. China}

\author{Cuihua Du}
\affiliation{College of Astronomy and Space Sciences, University of Chinese Academy of Sciences, Beijing 100049, P.R. China}

\author{Thomas Donlon II}
\affiliation{Department of Physics and Astronomy, University of Alabama in Huntsville, 301 North Sparkman Drive, Huntsville, AL 35816, USA}

\author{Mingji Deng}
\affiliation{College of Astronomy and Space Sciences, University of Chinese Academy of Sciences, Beijing 100049, P.R. China}



\begin{abstract}
Using elemental abundances for 1.26 million K giants in the LAMOST DR8 value-added catalog, we analyze the chemical abundances of the Virgo Overdensity (VOD) and Hercules-Aquila Cloud (HAC). 
We find two distinct chemical populations in both overdensities, which is in disagreement with the mainstream hypothesis that both overdensities are composed of materials from a single merger event, namely Gaia-Sausage-Enceladus (GSE). The two populations show different chemical trends: one exhibits low metallicities and high $\alpha$ abundances, and the other shows high metallicities and low  $\alpha$ abundances, which is associated with the recently discovered Nereus and Virgo Radial Merger (VRM) components in the local stellar halo, respectively. The Nereus component in these overdensities uniquely exhibits a decreasing trend in the [Fe/H]-[Mn/Fe] plane. Out of all observed Milky Way dwarf galaxies, this trend is only found in the Sculptor dwarf galaxy, which provides clues for the properties of Nereus progenitor. We also find that the velocity ellipse with high aniostropy parameters that is usually considered to be part of GSE are actually a mix of the two components. Both overdensities are well-mixed in kinematic spaces, confirming recent claims that the debris of merger pairs are kinematically indistinguishable in a recent simulation. We find that the velocity ellipses of the VRM stars in these overdensities have large inclination angles, which may be an indication of the merger time in simulated merger events. 

\end{abstract}

\keywords{Galaxy stellar halos(598) --- Galaxy kinematics(602) --- Galaxy dynamics(591)---Chemical abundances(224)}


\section{Introduction} \label{sec:intro}
According to the most widely accepted cosmological model ($\Lambda$-CDM), galaxies evolve through numerous accretion and merger events with smaller galaxies \citep{white1978,Blumenthal1984,Springel2005}. These merger events leave traces in phase space known as substructures \citep{Helmi&white1999,Newberg2009,H2018,B18,myeong_2019,koppelman_2019,Yuan_2020,Naidu_2020,Horta2021,Malhan_2022,kHYATI2024,liu24}. The most famous substructure, known as Gaia-Sausage-Enceladus (GSE), was discovered simultaneously and independently by \citet{B18} and \citet{H2018}, and is vividly characterized as a sausage shape in the $V_{r}-V_{\phi}$ plane. The progenitor of GSE was a massive satellite accreted $\sim 10$ Gyr ago \citep{B18,Gallart2019}, with an estimate of stellar mass $5 \times 10^{8} - 5 \times 10^{9} M_{\sun}$ \citep{Mackereth2019,Vincenzo2019}. The GSE debris is on highly radial orbit \citep{B18,Myeong18}, and is responsible for heating the proto-disk according to observational evidence (described as the ``Splash") and simulations \citep{Be2020,Fattahi19}. The corresponding substructures, observed overdensities \citep{Simion19,Perottoni,Ye24} the break of the stellar halo in the Milky Way (MW) \citep{Deason2011,Deason2018} all seem to support one single massive accretion event scenario (the GSE merger).

However, other works hypothesize that our Galaxy has gone through multiple mergers rather than one single major merger. \citet{Donlon19} originally claimed that the Virgo Overdensity is actually the result of a radial dwarf galaxy merger that they called the Virgo Radial Merger (VRM). The VRM is able to produce GSE-like debris and overdensities, but is instead thought to be only accreted $2 \sim 3$ Gyr ago. These authors also identified shell structures in the MW for the first time and found that shell structure disappears within 5 Gyr after the collision with the Galactic center in N-body simulations \citep{Donlon20}, which puts the impact time of the GSE into question. For the shell structures here, they are characterized by their distinctive appearance, resembling thin, elongated ``umbrella"-shaped clusters of stars, all located at a consistent distance from the Galactic center.
Later, \citet{Dollon23} (hereafter D23) chemically characterized the local stellar halo with APOGEE and GALAH data, and found four components using Gaussian Mixture Model (GMM), which they named VRM, Nereus, Cronus and Thamnos. Meanwhile, \citet{Kim21} presented evidence that multiple accretion events are required to explain observed structures, based on orbital inclinations and dynamical properties of GSE structures.
Recently, \citet{Horta2024} identified seven intermediate-age GSE stars, conjecturing that the last merger event could be younger than we think , or the local stellar halo is comprised of two satellites' debris (GSE progenitor with a younger companion). \citet{Dylan24} (hereinafter referred to as F24) used the IllustrisTNG50 simulation to identify stellar accretion histories in 98 Milky Way analogues, and found that the debris of a single merger and two mergers are indistinguishable in kinematic spaces unless also considering chemical abundances and star formation rates. This literature constrains the MW assembly history and put the ``one last major merger event" scenario into doubt.

\begin{figure*}
    \includegraphics[width=\textwidth]{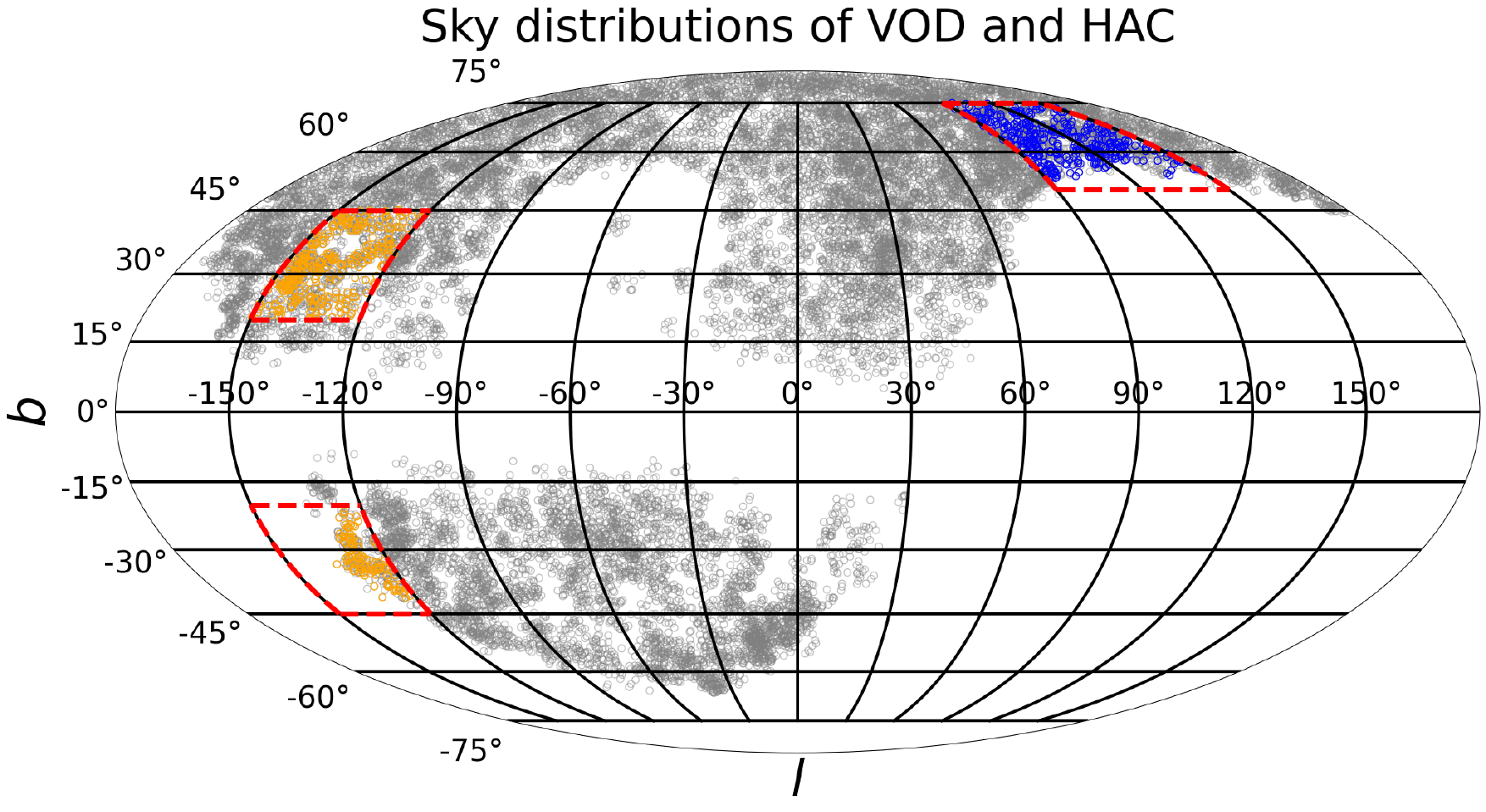} 
    \caption{The on-sky distributions of LAMOST K Giants in $l$ and $b$ coordinates. Grey dots represent the entire K giants sample from \citet{Zhang2023}, while blue dots and orange dots represent the VOD and HAC respectively. There are 366 stars in the VOD, 330 stars in HAC-N and 113 stars in HAC-S. }
    \label{Sky}
\end{figure*}

The Virgo Overdensity (VOD) and Hercules-Aquila Cloud (HAC) are two large and spatially diffuse cloud-like structures. The VOD was first identified and analyzed in RR Lyrae and  main-sequence stars \citep{Vivas2001,Newberg2002}. Later, \citet{Juric2008} identified a halo structure with sky coverage more than 1000 $\text{deg}^{2}$ in the direction of Virgo constellation. The HAC was first discovered by \citet{Belokurov2007} using main-sequence turnoff stars. The HAC extends across a heliocentric distance of $10 \sim 20$ kpc and is located at $l$ (25°, 60°) and $b$ ($
-$ 40°, 40°). Studies have shown the VOD and HAC are kinematically and chemically associated and perhaps share one common origin \citep{Simion19,Perottoni,Yan23}. However, when \citet{Perottoni} tried to chemically associate the VOD and HAC with GSE, chemical information was only available for a small number of stars (21 stars in the HAC and 22 stars in the VOD), making it difficult to draw any firm conclusions. Similarly, sample sizes in \citet{Yan23} were also small, and the corresponding chemical analyses lack iron-peak elements and are restricted in $\alpha$ abundances. Inspired by \citet{Dollon23} and \citet{Dylan24}, the VOD and HAC could contain debris from multiple mergers with indistinguishable kinematic properties, therefore chemical analyses of large samples for VOD and HAC are needed.

In this work, we use a sample of K giants with well-estimated elemental abundances to explore the chemical components hidden in the VOD and HAC, as well as kinematic differences of distinct populations. In section~\ref{data}, we describe the data and selection criteria. In section~\ref{result}, we show the chemical populations in the VOD and HAC after performing Gaussian mixture modeling (GMM). We also conduct an N-body simulation to better illustrate these results. In section~\ref{discussion}, we discuss the results. Finally, a comprehensive summary is given in section~\ref{summary}.

\begin{figure*}
    \includegraphics[width=\textwidth]{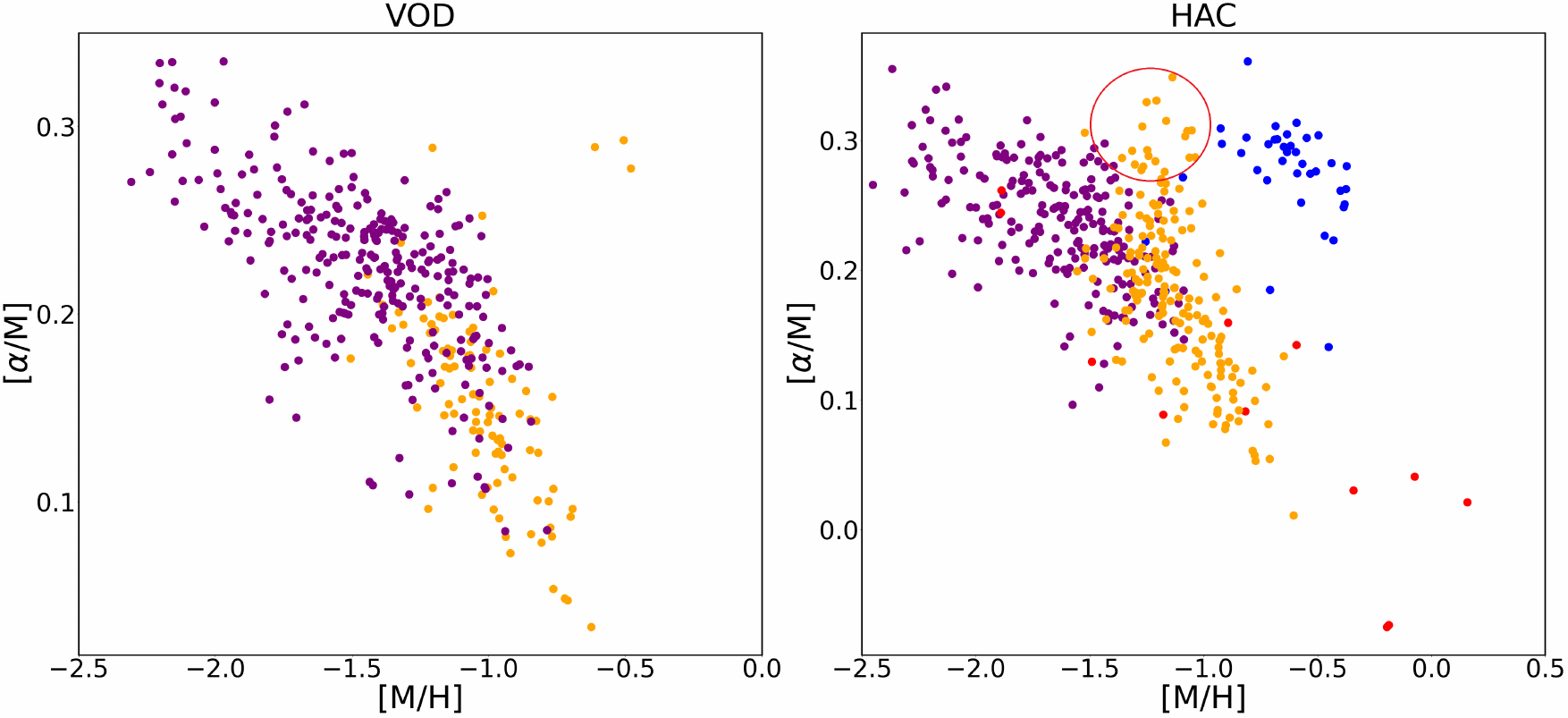}  
    \caption{GMM clustering results in the [M/H]-[$\alpha$/M] plane. Both the VOD and HAC have two components showing Nereus-like and VRM-like characteristics (purple dots and orange dots respectively). Additional two components in HAC are contamination from metal-rich halo (blue dots) and outliers recognized as a group by the GMM procedure (red dots), which are removed from the HAC sample for the rest of our analysis. Note that the HAC may be contaminated by Cronus stars as these stars are situated in the inner Galaxy (the red circle).  }
    \label{GMM}
\end{figure*}

\section{data} \label{data}
\subsection{Selection of VOD and HAC}

For the investigation of the VOD and HAC, we use a public catalog of distances for 19,544 K giants \citep{Zhang2023} drawn from the Large Sky Area Multi-Object Fiber Spectroscopic Telescope (LAMOST) DR 8 \citep{Cui2012}. Distances of K giants are derived based on estimates of absolute magnitudes in the SDSS $\it{r}$ band, and corresponding uncertainties are obtained from an Bayesian approach \citep{Xue2014} with a typical distance precision $\sim$ 11\%. For provided distances, we require \texttt{e\_Dist}/\texttt{Dist} $<$ 0.2 to ensure that these distances are accurate. We also cross-match the samples with \textit{Gaia} EDR 3 to obtain precise proper motions and uncertainties \citep{Gaia2016,Gaiaedr3}. As for elemental abundances, \citet{Li2022}
provided a value-added catalog for 1.2 million giants from LAMOST DR 8 with metallicity ([Fe/H] and [M/H]), total $\alpha$-abundance ([$\alpha$/M]) and nine elemental abundances (including C, N, O, Mg, Al, Si, Ca, Mn and Ni). These abundances are obtained using a neural network with mean absolute error between 0.02 $-$ 0.04 dex for most elements, thus no restriction on elemental abundance errors were imposed, as the majority are already below 0.2 dex. Finally we require line-of-sight velocity error $<$ 20 km/s for reliable velocities, which leaves 13,364 K giants in the sample. 

In order to select stars in the VOD and HAC regions, we follow the selection criteria in \citet{Perottoni}. The VOD is located at $l$ (270°, 330°) and $b$ (50°, 75°), while the HAC is divided into two parts ``HAC-N (north) and HAC-S (south)":  $l$ (30°, 60°) and $b$ (20°, 45°) for HAC-N; $l$ (30°, 60°) and $b$ (-45°, -20°) for HAC-S. For both overdensities, we only include stars with heliocentric distance between 10 kpc and 20 kpc. The final sample contains 366 stars in the VOD, 330 stars in HAC-N and 113 stars in HAC-S (see Figure ~\ref{Sky}).

\subsection{Orbital parameters}
Orbital parameters such as orbital energy ($E$), the z component of angular momentum ($L_{z}$),  maximum height above the mid-plane $Z_{max}$, and eccentricities ($e$) are calculated using \texttt{galpy} with the axisymmetric potential model \texttt{McMillan17} \citep{MC17}. According to the model's best-fit parameters,  we assume the Local Standard of Rest (LSR) velocity $V_{\text{LSR}}$ is 232 km/s and a radius distance of the Sun is 8.21 kpc, and the height above the mid-plane is 20.8 pc \citep{Bennet&Bovy}. For the solar motion, we adopt the values  $[U_{\sun}, V_{\sun}, W_{\sun}] = [11.1, 12.24, 7.25]$ km/s from \citet{schonrich}. We integrate the orbits forward for 10 Gyr to obtain above orbital parameters. The uncertainties for radial velocity $V_{r}$, azimuthal velocity $V_{\phi}$ and polar velocity $V_{\theta}$ in spherical coordinates are the standard errors after 100 Monte-Carlo sampling under orbital integration, where we assume Gaussian distributions for the uncertainties in proper motions ($\mu_\alpha$ and $\mu_\delta$), heliocentric distance ($d_{\sun}$) and line-of-sight velocity ($V_{los}$).

\begin{figure*}
    \includegraphics[width=0.5\textwidth]{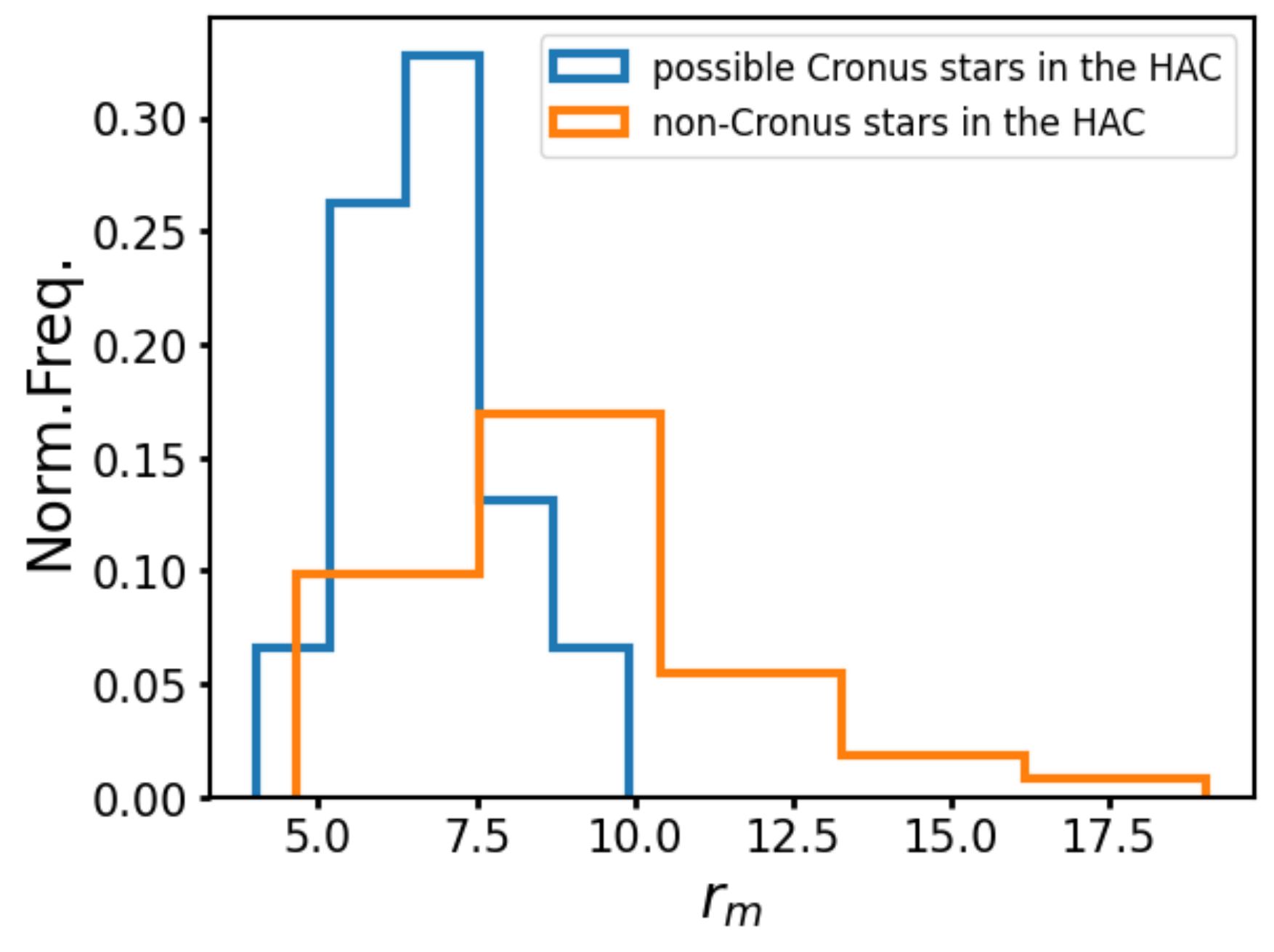}
    \centering
    \caption{The histogram of $r_m$ values for possible Cronus stars (10 stars in total with [Mg/Fe] $>$ 0.3) and non-Cronus stars in the HAC.}
    
    \label{hist}
\end{figure*}

\begin{figure*}
    \includegraphics[width=0.9\textwidth]{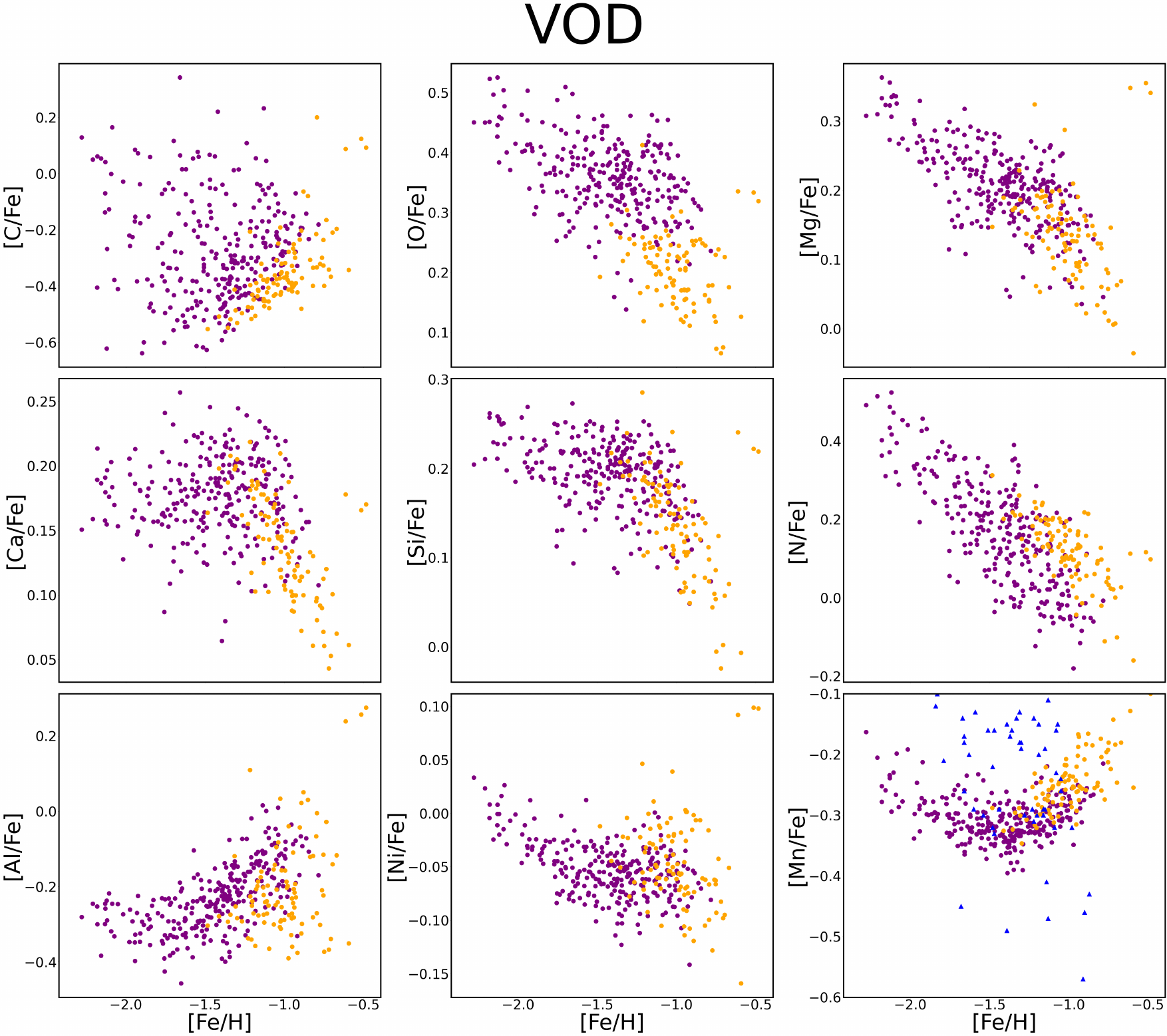} 
    \centering
    \caption{Chemical abundance distributions of nine elements in VOD. The purple dots and orange dots indicate Nereus-like and VRM-like components respectively. In this VOD samples, VOD is dominated by Nereus-like component with a weight of $\sim$ 73\%. The blue triangles in the [Fe/H]-[Mn/Fe] plane represent stars from the Sculptor dwarf galaxy \citep{North2012}, with Mn abundances measured using the Mn I 5407 Å line for clear comparison.}
    \label{VODmap}
\end{figure*}

\begin{figure*}
    \includegraphics[width=0.9\textwidth]{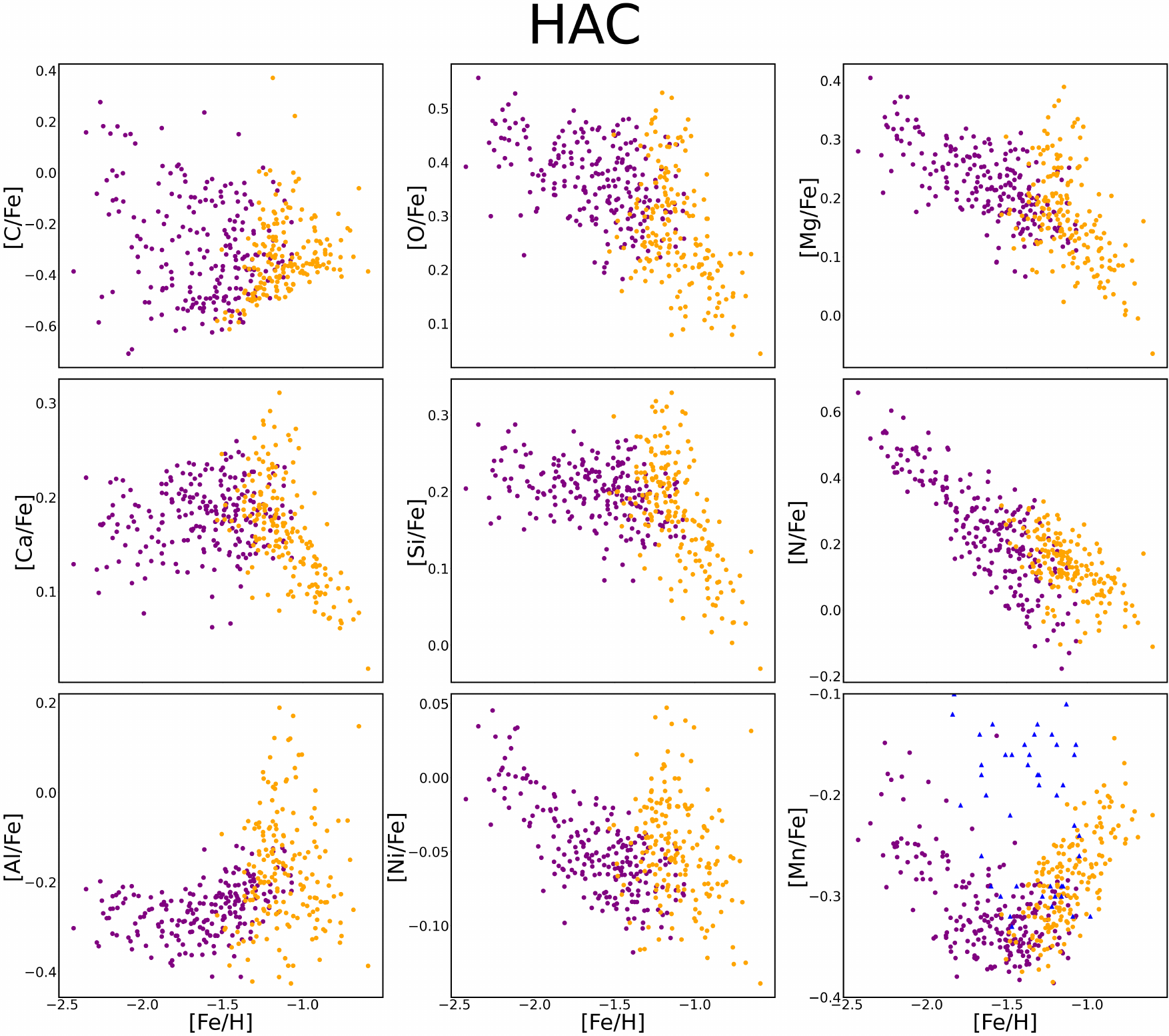}
    \centering
    \caption{Chemical abundance distributions of nine elements in the HAC. The purple dots and orange dots indicate Nereus-like and VRM-like components respectively. In this HAC samples, two components share similar weights ($\sim$ 54\% for Nereus-like component and $\sim$ 46\% for VRM-like component). The blue triangles in the [Fe/H]-[Mn/Fe] plane are same in Figure~\ref{VODmap}.}
    
    \label{HACmap}
\end{figure*}

\section{Results}\label{result}
\subsection{Chemical components in the VOD and HAC}
Following the method of D23, we apply the GMM from \texttt{scikit-learn} python package \citep{sklearn} to explore possible  
distinct chemical populations in the VOD and HAC. The GMM, as an unsupervised clustering algorithm, assumes that all data points are generated from a mixture of several Gaussian distributions with unknown parameters (including weights, means and standard invariance), combined linearly. These parameters can be estimated using the Expectation-Maximization (EM) algorithm \citep{Moon1996}, while the optimal number of GMM component is found by minimizing the Bayesian Information Criterion (BIC, Schwarz \citeyear{S1978}). GMM excels in handling overlapping clusters, assigning probabilistic memberships, and determining the optimal component number without requiring prior information through BIC method. However, one caveat is that GMM assumes that the data follows a Gaussian distribution, which may not always hold true, and it can be computationally intensive for large datasets or high-dimensional data. Therefore, compared to traditional supervised and hierarchical clustering, GMM is more intuitive, provides probabilistic cluster memberships, and is well-suited for this work.

The multi-dimensional quantities used by D23 are [Fe/H], [$\alpha$/Fe], [Na/Fe], [Al/Fe], $L_{z}$ and $\widetilde{E}$, where the last quantity is called ``pseudo-energy" and defined as follows:
\begin{equation}
    \widetilde{E} = E - \frac{V_{\phi}^2}{2}
\end{equation}
 $\widetilde{E}$ is used rather than the typical definition of energy to remove the correlation between $E$ and $L_{z}$, since both quantities are a function of $V_{\phi}$. However, the LAMOST data only contains information about $\alpha$ elements, meaning that we do not have access to [Na/Fe] abundances. As a result, we choose [Fe/H], [O/Fe], [Mg/Fe], [Si/Fe], [Ca/Fe], [Al/Fe], $L_{z}$ and $\widetilde{E}$ as inputs for the GMM analysis. In the VOD, the lowest BIC is obtained for two components, while the BIC is minimized for four components in the HAC. This result seems to support the ``multiple merger" scenario proposed by D23. This brings into question the idea that the VOD and HAC arise from the debris of a single merger, as is commonly believed to be the case for both the GSE and the VRM interpretations of these overdensities. This is not surprising, as the debris of pairs of mergers  are generally kinematically and spatially indistinguishable \citep{Dylan24}.

 The results after GMM clustering are shown in Figure~\ref{GMM}. Both the VOD and HAC show two similar components (purple dots and orange dots), with a clear separation at [M/H] $\sim -$ 1.3. The purple component exhibits Nereus-like characteristics (low metallicity and high $\alpha$-abundance) and orange component shows VRM-like characteristics (high metallicity and low $\alpha$-abundance). There are two additional components in the HAC data, which are removed for the rest of our analysis, as they appear to be related to structures unrelated to the overdensities. The blue component shows metal-rich halo-like characteristics, which could probably be the contamination from the metal-rich stellar halo \citep{Zhu2021}. The metal-rich stellar halo exhibits a metallicity range from $-1.0$ dex to $-0.4$ dex, with most stars displaying [$\alpha$/Fe] values between $0.2$ dex and $0.3$ dex. The red component is likely a group of outliers, as it only has 14 members. After removing these additional components, the HAC contains 390 stars.

\begin{table}[!htp]
    \centering
    \caption{Mean Elemental Abundances of Two Components in VOD and HAC} \label{one}
    \begin{tabular}{ccccc}
        \hline
        Abundance & VRM-like (VOD/HAC) & Nereus-like (VOD/HAC) & VRM (APOGEE/GALAH) & Nereus (APOGEE/GALAH) 
        \\
        \hline
        [Fe/H] &$-$ 1.00/$-$ 1.12 & $-$ 1.43/$-$ 1.62 & $-$ 1.14/$-$ 1.01 & $-$ 1.54/$-$ 1.43\\
        \hline
        [C/Fe] &$-$ 0.35/$-$ 0.33 & $-$ 0.27/$-$ 0.29 & $-$ 0.34/... & $-$ 0.28/...\\
        \hline
        [O/Fe] & 0.21/0.27 & 0.36/0.37 & 0.31/0.50 & 0.37/0.67    \\
        \hline
        [Mg/Fe] & 0.14/0.17 & 0.21/0.22 & 0.17/0.10 & 0.25/0.15    \\
        \hline
        [Ca/Fe] & 0.14/0.17 & 0.18/0.18 & 0.17/0.20 & 0.19/0.29    \\
        \hline
        [Si/Fe] & 0.14/0.17 & 0.19/0.20 & 0.19/0.12 & 0.23/0.28    \\
        \hline
        [N/Fe] & 0.12/0.13 & 0.17/0.23 & 0.14/... & 0.24/...    \\
        \hline
        [Al/Fe] &$-$ 0.20/$-$ 0.18 & $-$ 0.23/$-$ 0.26 & $-$ 0.21/$-$ 0.06 &$-$ 0.24/0.32    \\
        \hline
        [Ni/Fe] &$-$ 0.05/$-$ 0.05 & $-$ 0.05/$-$ 0.05 & $-$ 0.05/$-$ 0.16& $-$ 0.06/$-$ 0.09   \\
        \hline
        [Mn/Fe] &$-$ 0.25/$-$ 0.29 & $-$ 0.31/$-$ 0.31 & $-$ 0.34/$-$ 0.33 & $-$ 0.34/$-$ 0.30   \\
        \hline
        
    \end{tabular}
\tablecomments{The last two columns show the mean abundances of VRM and Nereus using APOGEE/GALAH data in \citet{Dollon23} for comparisons.}
\end{table}

 To investigate whether the two components we have identified in the VOD and the HAC actually correspond to the VRM and Nereus structures identified by D23, we provide chemical maps of the two components in the VOD and HAC respectively (see Figure~\ref{VODmap} and Figure~\ref{HACmap}). The mean values of elemental abundances in the two overdensities are summarized in Table~\ref{one}) to compare with the findings of D23. We find that the mean chemical abundances of the two components are comparable to the chemical properties of VRM and Nereus. Based on the overall similarity of the components we have identified and the VRM and Nereus components from D23, we are convinced that the two components in VOD and HAC are VRM and Nereus, thus we remove the affix ``-like" to directly call them VRM and Nereus hereinafter. However, it should be noted that there are some substantial differences between the chemical properties of these structures as measured by LAMOST compared to GALAH; these are perhaps due to observational bias, or differences in how the abundances are calculated in each survey (for example, these surveys use different spectral lines to measure each abundance). Specifically, LAMOST spectra have a broad and continuous wavelength coverage, while GALAH spectral coverage is segmented into four discrete arms. This segmentation may result in certain spectral lines falling within the gaps between arms, making them undetectable in GALAH data but observable in LAMOST spectra. Furthermore, as a high-precision spectroscopic survey, GALAH may detect weak spectral lines that are beyond the sensitivity limits of LAMOST.
 
 However, we also notice that there are some stars in the VRM component of HAC with larger $\alpha$-abundances, which might be contamination from Cronus stars (in red circles). The Cronus stars are located in the inner Galaxy, so they have small $r_{m}$ values (i.e, $(R_{ap} + R_{peri})/2$) and that is why they are missing in the VOD region. The Cronus member stars are relatively metal-rich ($-1.24\pm0.25$) and enhanced in $\alpha$ elements, with low energy and prograde properties, indicating an early time of accretion (e.g. see Figure 9 in D23). Alternatively, these stars could be contamination from ancient material left over from the proto-disk's spin up (also known as Aurora; Belokurov et al.\citeyear{Belokurov22}). We looked into the VRM stars in the HAC with [Mg/Fe] $>$ 0.3 and checked their $r_{m}$ values, as Cronus debris is predominantly at low Galactocentric radii. These stars showing $r_{m}$ values concentrated between 5 $-$ 8 kpc (see Figure ~\ref{hist}), possibly belong to Cronus stars with larger $r_{m}$ values (the values of Cronus components are 4.5 $\pm$ 1.5 kpc ). 
According to elemental distributions in Figure~\ref{VODmap} and Figure~\ref{HACmap}, we find that VRM stars in HAC have higher $\alpha$-abundances (even after removing the Cronus contamination). This possibly indicates that VRM stars in HAC were stripped earlier from their progenitor than those in VOD, due to the timescale for producing $\alpha$ elements in core-collapse (Type II, Ib and Ic) supernova \citep{Timmes1995,Kobayashi2006}.

Another surprising finding is that [Mn/Fe] exhibits a decreasing trend towards higher metallicity in Nereus, which is not discussed in D23. This rare trend is not seen in our Galaxy or corresponding evolution models \citep{Cescutti2008,Kobayashi2020,Eitner20} and from what we can tell, is only seen in the Sculptor dwarf galaxy and NGC 5139 \citep{North2012}, suggesting similarities between Sculptor dwarf galaxy and Nereus progenitor. \citet{boer12} used Red Giant Branch stars to study the chemical evolution and star formation rate of the Sculptor dwarf galaxy. In their work, the distribution of Sculptor stars in the [Fe/H]-[Mg/Fe] plane resembles that of Nereus stars, although the Sculptor stars exhibit enhanced magnesium ([Mg/Fe] $>$ 0.4 dex) at the metal-poor end.  Besides, the evolution of metallicity of Sculptor reaches [Fe/H] $\sim -$ 1.3 when t $\sim$ 3.5 Gyr after it was formed (see Figure 19 in Kobayashi et al.\citeyear{Kobayashi2020}). If the 
Nereus progenitor is indeed Sculptor-like, then it must have been accreted recently and could possibly have had a positive mean age gradient from the inner region to the outermost region \citep{Bett19} , since star formation can be quenched during mergers \citep{Ellison2022}. 

What needs stressing here is that this [Mn/Fe] downward trend characteristics might not be captured using GALAH data, because the GALAH survey goes down to 4718 Å \citep{GALAH} and only captures the strongest Mn I line at 4754 Å and 4823 Å both with a relative intensity of 40. By contrast, LAMOST covers a range from 3700 Å to 9000 Å, which helps capture the strongest line Mn I line at 4030 Å with a relative intensity of 1000 \citep{Cui2012}. \citet{Li2022} estimated the elemental abundances for giants from LAMOST DR8 between 4000 Å and 8500 Å, and the neural network method is able to capture Mn I line at 4030 Å due to its great intensity as well as weaker lines at 4033 Å and 4034 Å with relative intensity of 700 and 400 respectively. We consider the LAMOST Mn abundances to be reliable, and the downward trend observed in the [Mn/Fe]-[Fe/H] plane for the Nereus component is genuine.

We estimate the stellar mass of Nereus progenitor using the $M_{*}$([Fe/H], [Mg/Fe]) relation from \citet{Horta2021}. The authors fit the data to satellites of Milky Way-like galaxies in the L0025N075-RECAL simulation from \citet{M19}. The relation is as follows:
\begin{equation}
    \text{log}M_{*} = 10.28 + 2.18\langle[\text{Fe/H}]\rangle + 3.60\langle\text{[Mg/Fe]}\rangle - 0.30\langle\text{[Fe/H]}\rangle \times \langle\text{[Mg/Fe]}\rangle
\end{equation}
This relation produces a mass of $4.4 \times 10^7 M_{\sun}$ (HAC), $1.0 \times 10^ 8 M_{\sun}$ (VOD), $8.7 \times 10^7 M_{\sun}$ (APOGEE) and $5.8 \times 10^7 M_{\sun}$ (GALAH). During the fitting process, \citet{Horta2021} removed galaxies with fewer than 20 star particles (i.e., masses $\lesssim3.2\times10^7 M_{\sun}$) and those with $\langle$[Fe/H]$\rangle<-2.0$ dex to minimize the uncertainties. As a result, our estimates are considered reliable, as they fall within the specified constraints. The estimates already exceed the total mass of Sculptor, which is $\sim 3.1 \times 10^7 M_{\sun}$ \citep{Lokas2009}. This is curious, as we would expect Sculptor and Nereus to have similar masses due to their chemical similarity. However, if the extra mass in Nereus is due to the material from several additional minor mergers, this could drag up the mean [Mg/Fe] and result in a larger mass estimate. For the VRM progenitor, the estimated stellar masses are $3.2 \times 10^8 M_{\sun}$ (HAC), $4.4 \times 10^8 M_{\sun}$ (VOD), $2.9 \times 10^8 M_{\sun}$ (APOGEE) and  $2.9 \times 10^8 M_{\sun}$ (GALAH), still compatible with previous GSE progenitor $\sim 3 \times 10^8 M_{\sun}$ \citep{MB2020}. In D23, they argue that most selected ``GSE" stars are actually VRM stars and are a mix of VRM, Nereus and Thamnos (applying a cut on $\sqrt{J_{R}} >$ 30 km/s). As a result, it is difficult to identify Nereus without considering abundances, as Nereus has ten times fewer stars than the VRM and is therefore difficult to select in kinematic spaces. 

\begin{figure*}
    \includegraphics[width=\textwidth]{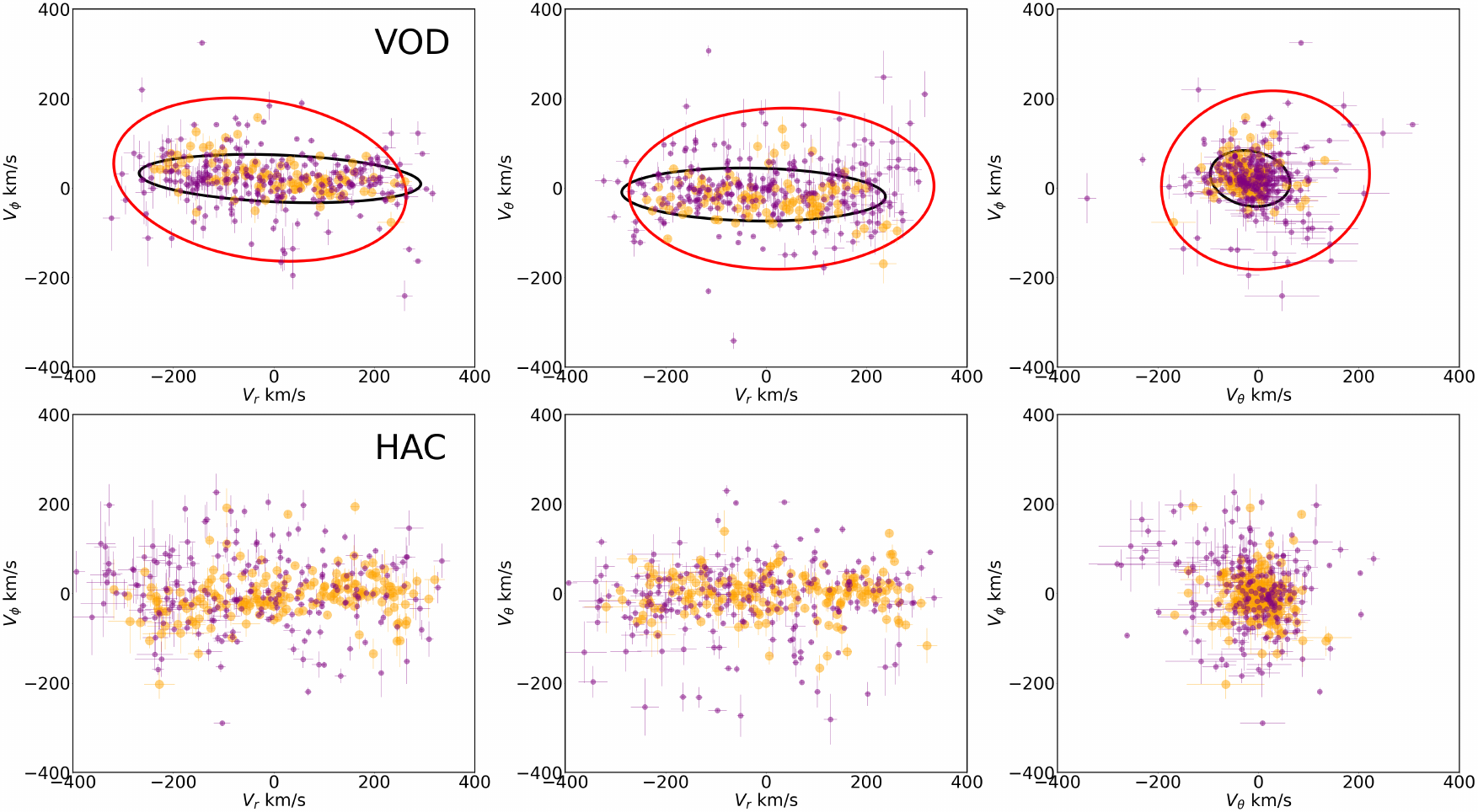} 
    \caption{Top row: velocity distributions of the VOD. Black and red ellipses represent highly anisotropic and less anisotropic components respectively. Notably, the highly anisotropic component is a mix of VRM and Nereus stars rather than one single population of GSE stars. This component contains most VRM stars and Nereus stars with high anisotropic parameters. Bottom row: velocity distributions of the HAC. The velocity ellipses are not given since there are more HAC stars with negative $V_{r}$, causing bias and indirect results for analysis. Purple dots and orange dots indicate Nereus and VRM components, respectively.}
    \label{Velocityellipse}
\end{figure*}

\subsection{Kinematics of VRM and Nereus in VOD and HAC}
Previous studies associate the VOD and HAC with the GSE hypothesis based on the fact that these structures all contain stars with similar velocity characteristics, which puts them on highly radial orbits. \citep{Simion19,Perottoni,Yan23,Ye24}. \citet{Simion19} used two multivariate Gaussians to fit the velocity data of VOD stars using the Extreme Deconvolution (XD) method \citep{Bovy2011} from the \texttt{astroML} package \citep{astroML12}. They attributed to the highly anisotropic part of the VOD to ``GSE" with a weight of $\sim$ 62\% and the less anisotropic part to the local halo. Here we take the same approach using two Gaussians but with VRM and Nereus indicated. From Figure~\ref{Velocityellipse}, the highly anisotropic component contains most of the VRM stars (which is reasonable because of its GSE-like characteristics) and anisotropic Nereus stars. The local halo component in \citet{Simion19} is probably made up of less anisotropic Nereus stars. We calculate the anisotropy parameters $\beta = 1 - (\sigma_{\phi} + \sigma_{\theta})/{2\sigma_{r}}$ \citep{BT2008} of two velocity components in the VOD, finding $\beta \sim$ $0.86_{-0.01}^{+0.01}$ with a weight of $\sim 71$\% (black) and $\beta \sim$ $0.71_{-0.01}^{+0.01}$ with a weight of $\sim$\% 29 (red), where the former weight is in agreement with $\sim 67$\% in \citet{Ye24}. 

\begin{figure*}
    \includegraphics[width=\textwidth]{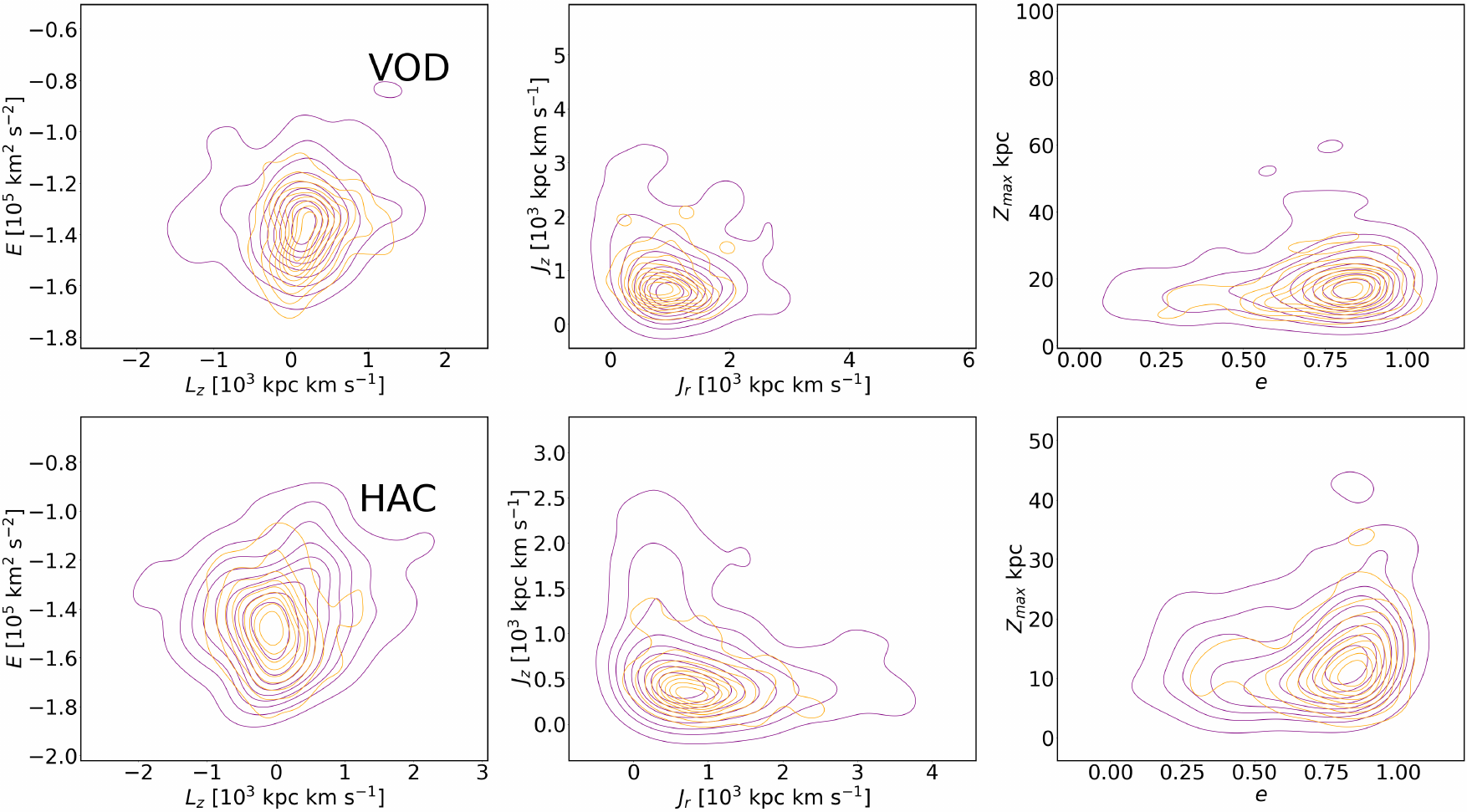} 
    \caption{Kinematic distributions of VRM and Nereus in both overdensities (purple contours for Nereus and orange contours for VRM).}
    \label{Kinematic}
\end{figure*}

F24 used simulations to find that merger pairs may yield kinematic distributions in which the galaxies comprising the merger are indistinguishable, occupying the same regions of these kinematic spaces. To verify this conclusion, we explore the kinematic distributions of VRM and Nereus in $E-L_{z}$, $J_{z}-J_{r}$ and $Z_{max}-e$ planes in Figure~\ref{Kinematic}. As the figure illustrates, two components in both overdensities are well-mixed in three planes. However, Nereus stars tend to have higher energy than VRM stars with mean differences $\sim$ 5,974 km$^2$ s$^{-2}$ (VOD) and $\sim$ 5,453 km$^2$ s$^{-2}$ (HAC), while D23 noted the opposite trend. There are also Nereus stars with higher $J_{z}$ values and lower eccentricities than those of VRM stars (see Figure~\ref{e} for better comparisons), possibly indicating the inclination differences of their progenitors merging with MW \citep{Lane2022}. To further investigate kinematic differences behind the two components, we took velocities, actions as well as orbital frequencies as input using \texttt{RandomForestClassifier} from sckit-learn package and split the data to train them \citep{sklearn}. Unfortunately and unsurprisingly, the precision is just a little above 0.5 and all input parameters share similar weights, which means the classifier failed to label untrained data. As a result of this analysis, we confirm that if Nereus and VRM had different progenitors, they are now indistinguishable in kinematic spaces. This is consistent with the finding of F24 that pair (or even multiple) mergers overlap in kinematic spaces, and as a result any analysis of Nereus and VRM must use elemental abundances in order to identify these structures.

\subsection{Simulation}
We also notice that there are clearly velocity ellipse tilts seen in Figure~\ref{Velocityellipse} and in \citet{Ye24} (see their Figure 6 and Figure 7 for details), which could be due to incomplete phase-mixing since VRM is a late merger event. To better understand the relationship between the angles of inclination and the time of merger, we use the GIZMO code \citep{gizmo} to re-construct an N-body simulation of the VRM event based the model from \citet{Donlon20}. The Milky Way and the VRM progenitor are set with a distance of 30 kpc with an inclined angle 30$^\circ$ from the MW panel, and the VRM progenitor has no initial velocities in this simulation for simplicity. The parameters of the MW model are adopted from \citet{Naidu2021} and \citet{Wang2022}, the VRM progenitor is based on \citet{Donlon20}, and the rotation curve of the MW in our simulation is fitted to observational data from \citet{Zhou2023}. For the MW, the dark matter and the bulge are in Hernquist \citep{Hernquist} and Einasto \citep{Einasto} profiles, respectively. And the disk model follows an exponential + sech-z profile. For the VRM progenitor, the stellar mass distribution adopts Plummer profile \citep{Plummer}. The detailed parameters are shown in Table ~\ref{table2}.

\begin{figure*}
    \includegraphics[width=\textwidth]{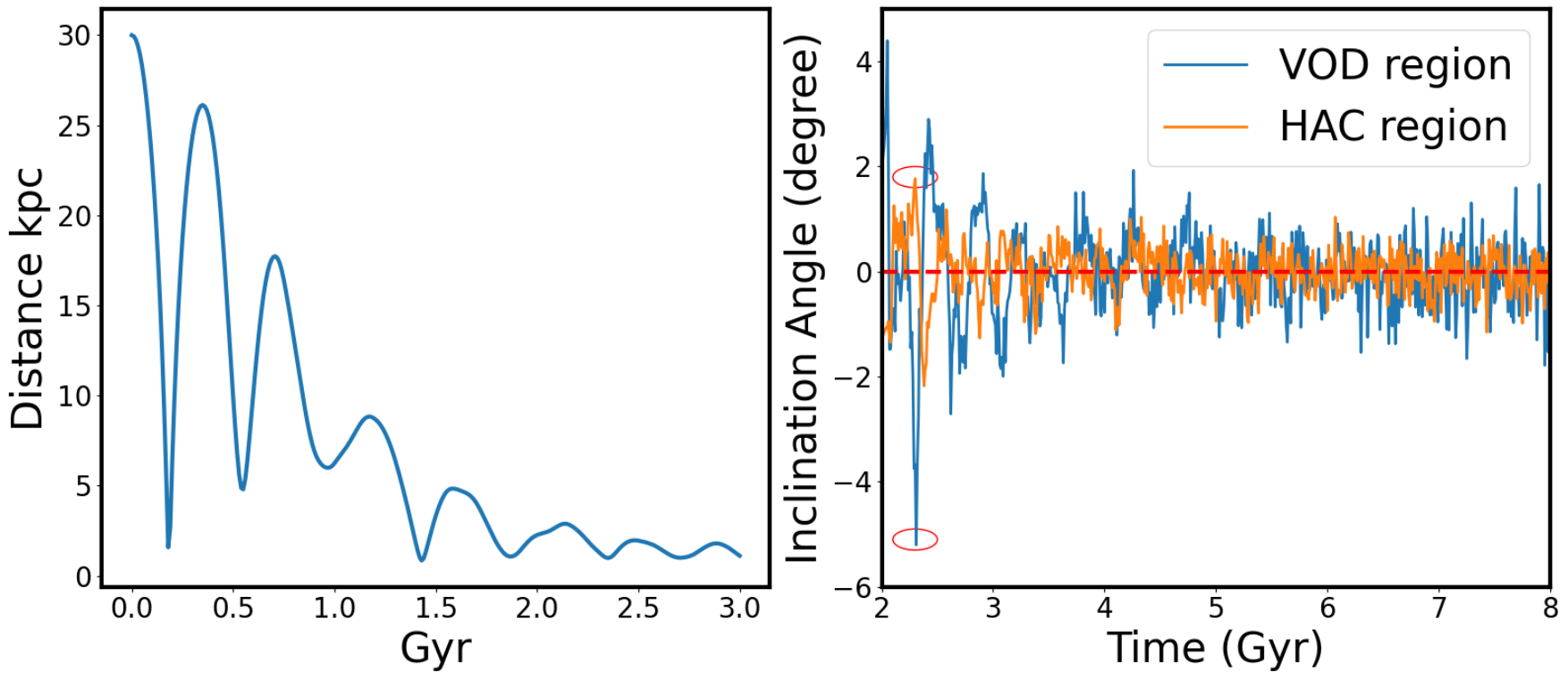} 
    \caption{Left panel: the distance between the centroid of VRM progenitor and the Milky Way, where the merger is almost complete at $\sim$1.5 Gyr. Right panel: the inclination angles as a function of time. The red circles at $\sim$ 2.3 Gyr indicate the angles resemble the observed situation (in terms of the ratio).}
    \label{simulation}
\end{figure*}

For the spatial selection for VOD and HAC, we choose the overdensity in the northern hemisphere as VOD and the opposite overdensity as HAC at the simulation snapshot at 3 Gyr. This selection is somewhat arbitrary, because the VOD and HAC regions move in space throughout the simulation, making it unlikely for them to align with the currently observed positions at any given time. As long as we are considering the over-dense regions of the simulated halo, this selection will be sufficient to qualitatively analyze the simulation. As seen in the left panel of Figure ~\ref{simulation} , the VRM progenitor experiences its second pericentric passage $\sim$ 0.5 Gyr into the simulation, and the merger is almost complete around 1.5 Gyr. We traced the IDs of selected VOD and HAC stars to calculate their inclination angles in $V_{R}-V_{\phi}$ plane as a function of simulation time between 2 and 8 Gyr (see the right panel of Figure~\ref{simulation}). It can be observed that the changes in inclination angles over time are a persistent characteristic, and these angles decrease in amplitude until about 4 Gyr after the simulation, after which time they maintain the same amplitude. This is about the same amount of the time that it takes shell structure to disappear \citep{Donlon20,Donlon24}. The inclination angles of VOD and HAC do not consistently exhibit opposite signs; for nearly half of the time, they have opposite signs, while at other times, they share the same sign. 

The inclination angles of the VRM component in VOD and HAC in our samples are $-8.59$ ° and $4.00$ °, respectively (with velocity errors considered). Although the magnitudes of the angles do not appear to align well with the simulation results, the simulation indicates that the angles could be large at earlier times, while the absolute values of the angles are below 2° after 3 Gyr. A similar situation is observed in our simulation, where at $\sim$ 2.3 Gyr, the angles of VOD and HAC are $-5$° and 2°, respectively (noted by the red circles). Therefore, we assume that the large angles can only occur at early times, which indicates that VOD and HAC could be a mix of debris from younger mergers. Whether the inclination angles really indicate the merger time needs detailed analysis in future work.

\begin{table}[!htp]
    \centering
    \caption{Parameters for Milky Way and VRM models} 
    \label{table2}
    \begin{tabular}{|c|c|c|} 
        \hline
        \ Parameter & Milky Way Model& VRM Model \\ 
        \hline
        Mass 
        & total: $5 \times 10^{11} M_{\sun}$ & total: $1 \times 10^{9} M_{\sun}$ \\ 
        & dark matter:  $4.487 \times 10^{11} M_{\sun}$ (Hernquist profile) & stellar mass: $1 \times 10^{9} M_{\sun}$ (Plummer profile)\\ 
        & stellar disk: $3.58 \times 10^{10} M_{\sun}$ (exponential + sech-z profile) & \\ 
        & bulge: $1.55 \times 10^{10} M_{\sun}$ (Einasto profile) & \\ 
        \hline
        Scale 
        & stellar scale length: $2.4$ kpc & stellar scale length: $3$ kpc \\ 
        & dark matter scale:  $42$ kpc & \\ 
        & bulge scale: $1.5$ kpc & \\ 
        \hline
    \end{tabular}
    \tablecomments{The parameters for the Milky Way model is primarily based on \citet{Naidu2021} and \citet{Wang2022} with some modifications according to \citet{Donlon20}. The circular velocity curve of the Milky Way model is well-matched with the data from \citet{Zhou2023}.}
\end{table}

\section{Discussion}\label{discussion}
\begin{figure}[h] 
    \centering 
    \begin{minipage}[t]{0.46\textwidth} 
        \includegraphics[width=\textwidth]{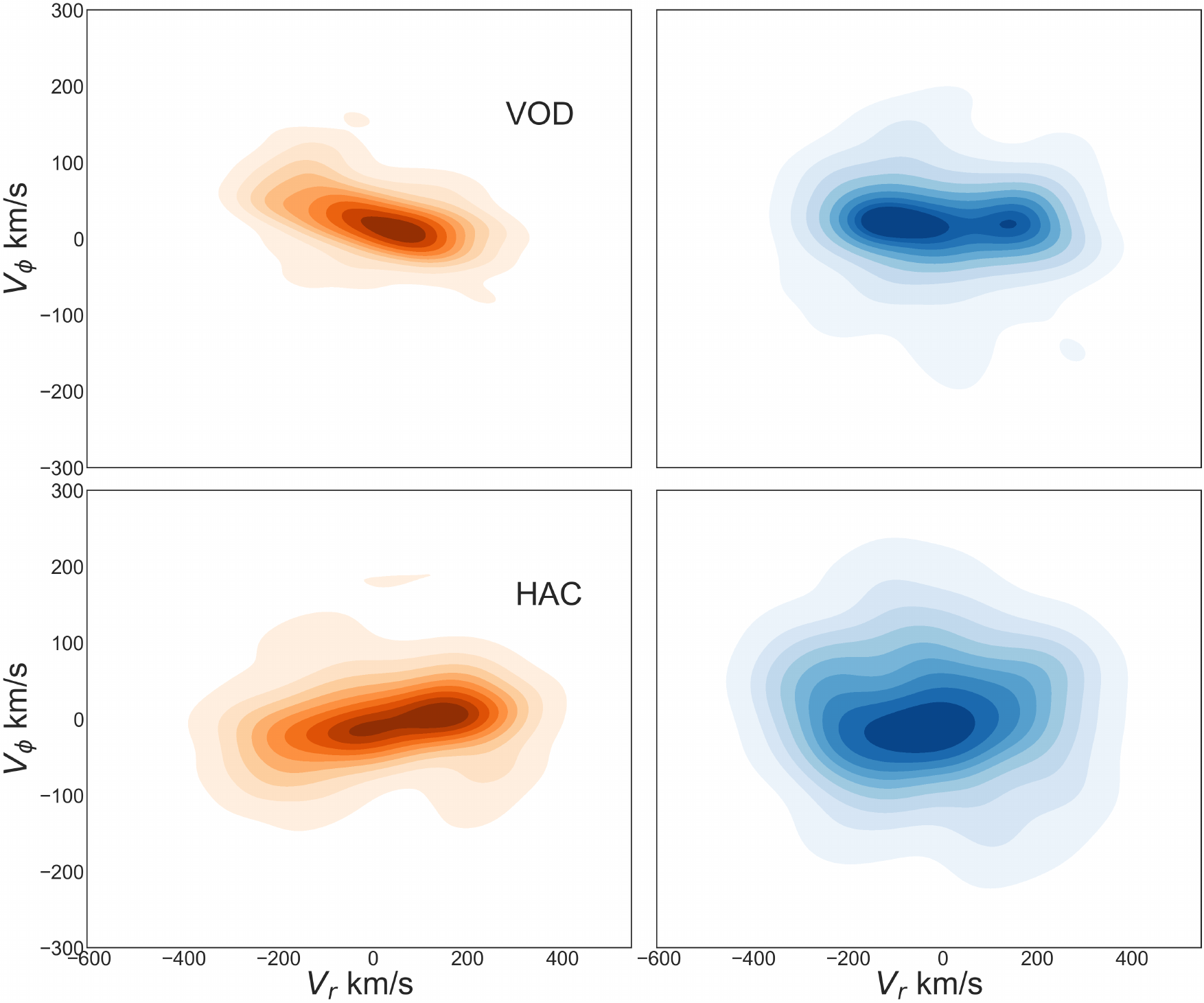} 
        \caption{Top row: KDE maps for the VOD in $V_{r}-V_{\phi}$ plane. Bottom row: KDE maps for the HAC in $V_{r}-V_{\phi}$ plane. Orange and blue colors represent VRM and Nereus components respectively. Note the significant tilt of the velocity distribution for VRM stars.}
        \label{vr}
    \end{minipage}
    \hfill 
    \begin{minipage}[t]{0.46\textwidth} 
        \includegraphics[width=\textwidth]{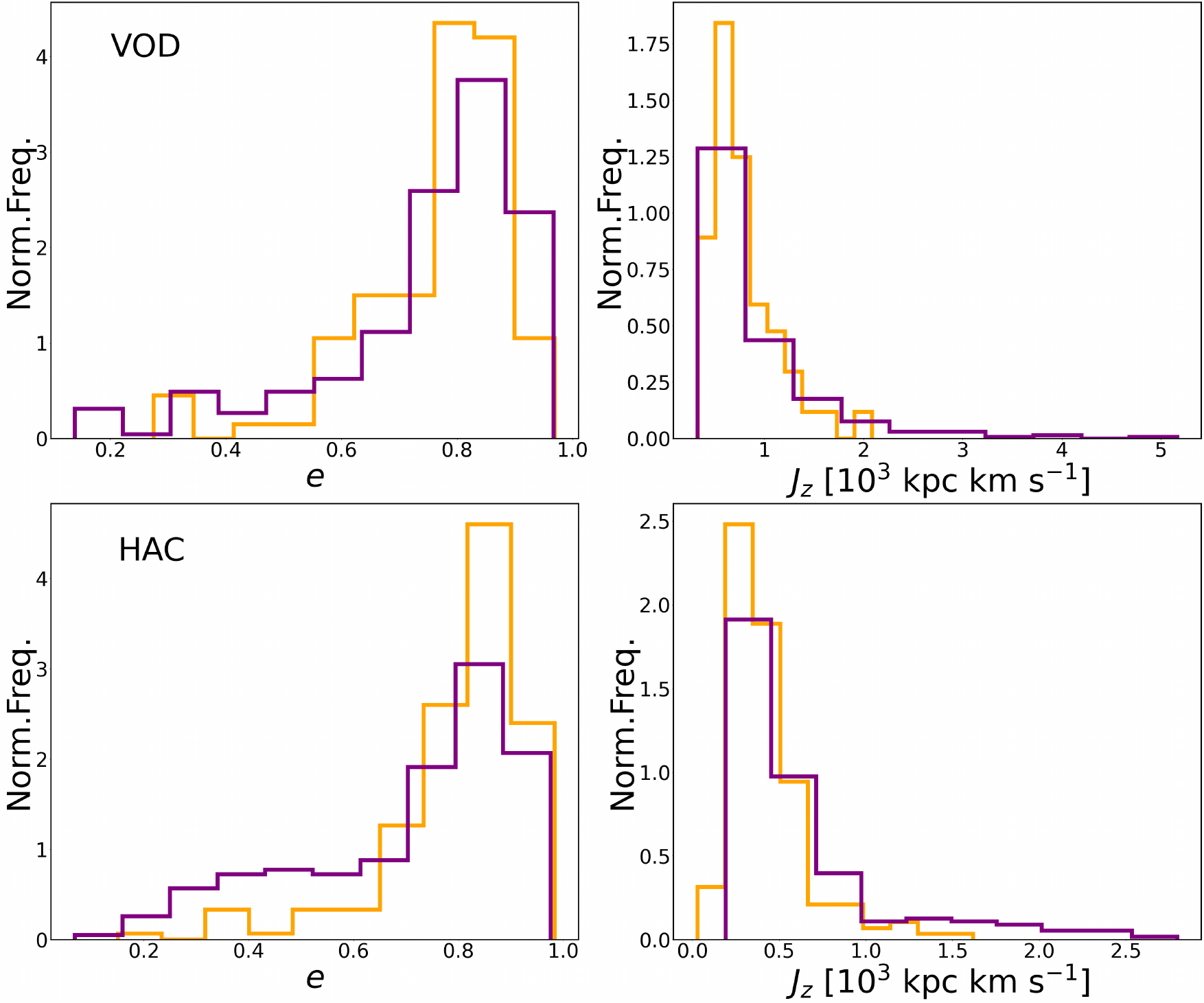} 
        \caption{Top row: histograms of $e$ and $J_{z}$ in the VOD. Bottom row: histograms of $e$ and $J_{z}$ in the HAC (purple for Nereus and orange for VRM). Notably, Nereus has more low $e$ stars and a long tail towards high $J_{z}$ values (more than 2,000 kpc km s$^{-1}$), which is not seen in GSE-like merger \citep{A2022} and indicates a different origin of Nereus.}
        \label{e} 
    \end{minipage}
\end{figure}

 We have shown that VRM and Nereus are indistinguishable in kinematic spaces, but have distinct chemical characteristics. However, their kinematic differences can still be traced by star motions in the condensed region. We re-plot the $V_{r}-V_{\phi}$ plane for these regions using Kernel Density Estimate (KDE; Gaussian kernel is chosen) in Figure~\ref{vr}. We notice that most stars in main density center of Nereus have negative $V_{r}$ values, while those of VRM have positive values. This means that the center region of Nereus is moving away from the Galactic center and that of VRM is moving towards the Galactic center, which provides a constraint on the properties of the VRM and Nereus progenitors. This is an interesting and important point -- if the GSE was very old and the entire VOD and HAC came from a single merger event, then one would expect there to be an equal number of stars with positive and negative $V_r$, which is clearly not what we see here. This is shown even more clearly by the fact that chemistry appears to be linked to $V_r$ -- so the main bulk of Nereus and VRM are reaching the VRM and HAC at different times (presumably because they fell in with slightly different energy). Although it should be pointed out that this same effect could happen if a single large dwarf with a metallicity gradient fell in (see Figure 3 of D23). Notably, although VRM and Nereus show differences in their velocity distributions, this does not imply kinematic distinguishability. As shown in Figure~\ref{Kinematic}, their kinematic parameters are largely overlapping. Nevertheless, subtle differences might still provide clues to variations in their progenitor properties, even if kinematically inseparable.

Another noteworthy overdensity is Eridanus–Phoenix Overdensity (EriPhe overdensity) is an access of main-sequence turnoff stars in the direction of the constellations of Eridanus discovered by the Dark Energy Survey \citep{Li2016}. This overdensity is centered around ($l$, $b$) $\sim$ (285°, -60°), spanning at least 30° in longitude and 10° in latitude with $d_{\sun} \sim$ 16 kpc. The authors associate EriPhe overdensity with VOD and HAC according to a polar orbit, which is similar to the Vast Polar Structure (VPOS) plane \citep{P12,P15}. Both \citet{Simion19} and \citet{Yan23} integrated the orbits of VOD and HAC over past 8 Gyr, but no clear density center is seen in EriPhe region in $l-b$ plane. Therefore, we are unable to support the claim that EriPhe is related to the VOD and HAC, and a more thorough exploration of the origins of EriPhe will require additional data that may become available in the future.

\section{Summary}\label{summary}
Based on K giants elemental abundances, we apply GMM to analyze chemical components of the VOD and HAC using a relatively large sample size. Our conclusions are summarized as follows: 

(1) We find that there are two distinct components in the VOD and HAC, which correspond to the VRM and Nereus structures in D23. This challenges the ``last major merger event'' (i.e. GSE) scenario. 

(2) Nereus shows a rare decreasing trend in the [Mn/Fe]-[Fe/H] plane, which is only seen in Sculptor among observed dwarf galaxies. This may provide constraints as to the nature of Nereus, such as its star formation rate. However, whether this trend holds true requires further confirmation with additional data from upcoming high-resolution surveys.

(3) The highly anisotropic part of the VOD is actually a mix of Nereus and VRM, rather than a single GSE component. 

(4) Nereus and VRM are kinematically indistinguishable, consistent with F24's findings. This suggests that stellar halo structures identified purely through kinematics may harbor multiple progenitors detectable via elemental abundances. Although their kinematic overlap persists, the distinct $V_r$ trends of their density centers (predominantly positive for VRM and negative for Nereus) could encode progenitor-specific merger signatures. However, these differences likely reflect intrinsic compositional properties rather than kinematic divergence, particularly given the limitations of machine learning in separating these components.

(5) Our simulation suggests that large inclination angles of velocity ellipse in $V_{R}-V_{\phi}$ plane could correspond to more recent mergers, although this property remains to be examined in more detail in future work.

These discoveries provide additional evidence that the GSE material is actually comprised of debris from multiple merger events, which provides valuable insight about the assembly history of the Milky Way and its evolution.

\section{ACKNOWLEDGENMENTS}
This work was supported by National Key R\&D Program of China No. 2024YFA1611900, and the National Natural Science Foundation of China (NSFC Nos. 11973042, 11973052). This work is based on data acquired through the Guoshoujing Telescope. Guoshoujing Telescope (the Large Sky Area Multi-Object Fiber Spectroscopic Telescope LAMOST) is a National Major Scientific Project built by the Chinese Academy of Sciences. Funding for the project has been provided by the National Development and Reform Commission. LAMOST is operated and managed by the National Astronomical Observatories, Chinese Academy of Sciences.

\bibliography{liu}{}
\bibliographystyle{aasjournal}



\end{document}